ARTICLE

# Liquid Crystalline Assembly of Collagen for Deterministic Alignment and Spread of Human Schwann Cells

Homa Ghaiedi[a], Luis Carlos Pinzon Herrera[b], Saja Alshafeay[c], Leonard Harris[c], Jorge Almodovar[b] and Karthik Nayani[a*]

[a] Ralph E. Martin Department of Chemical Engineering, University of Arkansas, Fayetteville, AR, 72701, USA

[b] Department of Chemical, Biochemical, and Environmental Engineering, University of Maryland Baltimore County, Baltimore, MD, 21250, USA

[c] Department of Biomedical Engineering, University of Arkansas, Fayetteville, AR, 72701, USA

Collagen is a key component of the extracellular matrix and well-oriented domains of collagen are relevant for mimicking the local cell environment in vitro. While there has been significant attention directed towards the alignment of collagen, formation of large-scale oriented domains remains a key challenge. Type I collagen self-assembles to form liquid crystalline (LC) mesophases in acidic conditions at concentrations above 100 mg/ml. The LC mesophase provides an efficient platform for large-scale alignment and patterning of collagen coated substrates. However, there exist challenges related to solubilizing and processing of collagen at such high concentrations in order to replicate the native extra cellular matrix (ECM). In this contribution, we report on centimeter-scale alignment in collagen-coated glass substrates using solutions that are well below the LC-forming concentrations. Importantly, we are also able to extend this method to create a mimic of the native ECM via macroscopic 3-D collagen hydrogels with programmed anisotropy within them. We explain the formation of these uniform domains via shear-induced and magnetically-induced liquid crystallinity of the collagen solutions. We show that the orientation, spreading and aspect ratio of Human Schwann Cells (HSCs) all are strongly coupled with the alignment of the collagen substrate/hydrogel. We use a simple Metroplis-based model to reveal that a critical magnetic field strength exists for a given concentration of collagen, exceeding which, macroscopic alignment is permissible- enabling guidance for future studies on alignment of collagen at high concentrations.

## 1. Introduction

Regeneration of damaged peripheral nerve, due to trauma or neurological disease, impacts approximately 20 million people in the US and corresponds to an annual healthcare expense of approximately $150 billion [1, 2]. Nerve guide conduits (NGCs) are becoming more prominent as a substitution for autologous allografts as they eliminate the need for multiple surgeries to harvest the autologous tissue. NGCs have shown great promise in the regeneration of short nerve defects; however, they fail in the regeneration of large nerve defects due to the lack of guidance cues which can deterministically proliferate nerve cells. An approach to address this challenge is by using templated substrates as NGCs that are capable of imparting guidance cues for the growth and proliferation of Human Schwann Cells (HSCs). We wager that LCs (fluids with orientational order) provide an efficient platform to design NGCs incorporating well-defined orientational cues. Our hypothesis is inspired by findings that collagen self-assembles within the body to form a myriad of LC phases. From a fundamental perspective, our study also aims to probe the coupling of substrate anisotropy and reconfiguration of cells in response to mechanical cues. Our work advances biologically inspired design of LC-based hydrogels with internal ordering and properties that can be finely tuned, enabling directional growth of cells.

HSCs play a vital role in the regeneration of peripheral nerve through secretion of various proteins that affect the behavior of glial cells in formation of cellular channels that are responsible for guiding axons to the correct targets[3]. The orientational fidelity of HSCs promotes axonal outgrowth, which in turn controls their orientation [4, 5]. Therefore, understanding the fundamental principles by which the directional behavior of

[a.] Department of Chemical Engineering, University of Arkansas, 3202 Bell Engineering Centre, Fayetteville, AR 72701, USA
*. Email: knayani@uark.edu

HSCs can be mediated is necessary. Several reports have focused on the effects of spatial configuration of the extracellular matrix (ECM) on the alignment of various biological cells including HSCs [6-10]. There are only a few studies associated with the role of alignment/orientation of the substrate and their influence on HSCs [11]. Moreover, these studies are primarily limited to 2D environments that are incapable of mimicking the native ECM. 2D environments are insufficient to replicate the native environment as, the growth pattern of cells is highly influenced by the topography of ECM [12-14]. Changes in the morphology of cells upon exposure to constraints of a 2D environment affect the proliferation rate, signaling and secretion [15]. Alteration of phenotypic properties and loss of polarity due to the poor interaction with ECM in 2D environments can also to an insufficient response to apoptosis [16]. The intracellular interaction in the native environment also has enormous impact on the epithelial collagen production and regulation of intercellular events [17]. Therefore, it is important to have a 3D environment that can imitate the native ECM if any extracellular event is to be accurately investigated.

The spatial arrangement of ECM plays a vital role in controlling the behavior of biological cells [18-20]. Collagen as a major component of native ECM has shown to self-assemble into chiral nematic LC structures at concentrations above 100 mg/ml, enabling cells to anchor to and elongate along the long axis of the composed fibrils [21-25, 26, 27]. Additionally, the alignment of LCs can be triggered by several stimuli including flow, electric or magnetic field, and underlying alignment layers [28-31]. In this study, we aim to investigate how a mimic of native-ECM using well-aligned collagen networks can affect the directional orientation and spread of HSCs. To achieve this, it is critical to develop facile methods to prepare films and hydrogels constituted of anisotropically arranged collagen and subsequently, to investigate the orientational behavior of HSCs cultured in such environments.

Here, we harness the inherent anisotropic viscoelastic properties and diamagnetic susceptivity of collagen to create novel platforms composed of long-range unidirectionally aligned cholesteric bands (at concentrations well below the reported critical concentration of collagen to create LC phase), both in films and hydrogels and study the effect of anisotropy n the alignment and spreading of HSCs. While the alignment of HSCs along the direction of collagen orientation is intuitively expected, our results also show that, remarkably, the aspect ratio of HSC is a function of the alignment of collagen.

## 2. Results and Discussion

Understanding the process driving the formation of liquid crystalline collagen enables greater control over the configuration of these assemblies, ultimately enabling the design of scaffolds that are more fitting for specific applications. Collagen Type I is composed of three polypeptide chains wound together in a left-handed triple helix (depicted in Fig. 1A). Collagen self-assembles into fibrils with diameter of 1 nm and length of 300 nm. At low concentrations, collagen exists in an isotropic phase characterized by the absence of a preferred orientation of the constituent collagen fibrils. Upon increasing the concentration above 100 mg/ml in acidic environments (pH = 2.5-3), however, collagen demonstrates liquid crystalline behaviour, as evidenced by the formation of oriented cholesteric bands composed of fibrils arranged in a hexagonal packing pattern [23, 32]. The directional preference seen in liquid crystals arises from the collective behaviour of macromolecular assemblies comprised of asymmetrical molecules. This is consistent with an Onsager-like description wherein the LC-phase arises from interplay between orientational entropy and excluded volume [33]. The system seeks to balance the need to maximize the overall entropy by increasing its positional entropy and sacrificing its orientational freedom, leading to the formation of the LC phase.

### 2.1. Preparation and characterization of aligned collagen films with microgrooves and shear

Polyimide (PI) coated glass surfaces that are rubbed uniaxially to create microgrooves are a well-established method to create well-defined alignment in LC films[34, 35]. In this vein, we create aligned collagen films using a combination of PI-coated glass substrates and shear force. The motivation to use shear-driven alignment stems from the objective to use concentrations of collagen that are well-below LC forming concentrations [36]. Shear-driven alignment was achieved using a doctor blade to deposit collagen solutions (with concentrations as low as 10 mg/mL) on glass substrates. Prior to the deposition process, microgrooves were formed on PI-coated microscope glasses, by rubbing the glass in a unidirectional manner with a velvet cloth. Under cross-polarized illumination, a planar arrangement of birefringent cholesteric bands was observed in the collagen films with concentration of 10 mg/mL (Fig. 1B). In contrast, Fig. 1C shows the absence of such optical features for 10 mg/mL collagen solution spaced between a cover slip and a microscope slide under cross-polarized light. The complete extinction of light in this image indicates the random orientation of fibrils in the absence of shear force, thus highlighting the important role of shear force in the formation of aligned cholesteric bands. Fig. 1D depicts the trajectory of fibrillar orientation along the major axis of cholesteric bands.

The degree of alignment of cholesteric bands can be quantified using the order parameter (equation 1). As a marker for the alignment strength of the collagen films, we quantify the order parameter (S) of the cholesteric bands by evaluating the alignment of individual cholesteric bands. In our experiments, the order parameter for unidirectionally aligned cholesteric bands over an area of 6 cm² was determined to be as high as 0.85, suggesting a high degree of alignment[31, 37, 38].

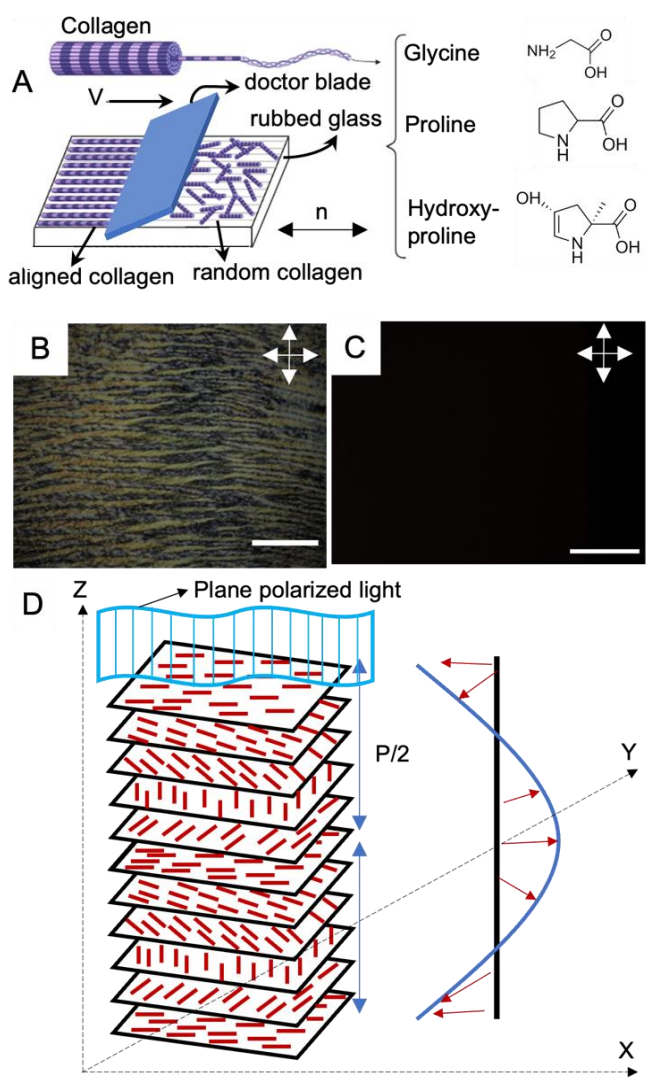

**Figure 1.** A) Schematic demonstration of shear-driven alignment of collagen on rubbed PI coated substrate. Collagen solution was deposited using a doctor blade upon which a film composed of unidirectionally aligned cholesteric bands formed spontaneously. (B) Cross-polarized optical micrograph of LC collagen (10mg/ml in aqueous acetic acid (pH 2.5)) film, which has been coated on a PI-coated glass substrate that was previously rubbed to induce directional alignment. The characteristic pattern of alternating light and dark bands, known as cross-striations, is clearly visible in the cholesteric bands.(C) Cross-polarized optical micrograph of collagen solution (with the same concentration and pH 2.5) between a coverslip and a microscope glass spaced by 150µm, serving as a control to evaluate the effect of shear on the alignment of cholesteric bands. (The scale bar represents 100µm in both B and C.) (D) Schematic of arrangement of collagen molecules exhibiting a left-handed twist across parallel planes. The distance between layers, denoted as P is equal to the full rotation of the long axes at 360°, and it is known as the pitch. The double-sided arrows indicate the direction of analyzer and polarizer.

$$S = \langle \tfrac{3}{2} \cos^2\theta - \tfrac{1}{2} \rangle \quad (1)$$

The pitch length, denoting the extent over which a complete rotation of collagen fibrils occurs, was quantified using ImageJ software. Two parallel lines, aligned with the orientation of consecutive dark regions, were delineated, and the perpendicular separation between these two parallel lines was defined as the pitch. The determined pitch value is 11.95 µm, with a standard deviation of 0.556 µm.

To further quantify the liquid crystalline nature of the aligned collagen films, observations were conducted using cross-polarized microscopy integrated with a rotating stage. The stage was rotated through small incremental rotations, and the average grayscale intensity of the resulting micrographs was quantified. The alterations in interference colour and the changes to the grayscale intensity are classical markers of unidirectional alignment of LCs. As shown in Fig. 2, the unidirectional alignment of cholesteric bands is evident, substantiated by both the cosine trend (dashed line which is the fit) and the four-fold symmetry plot of the distribution of the grayscale intensity versus the rotation angle.

### 2.2. Characterization of the behaviour of HSCs cultured on aligned collagen films

Aligned collagen films were placed in 6-well culture plates and subjected to a culture medium comprised of HSCs and a low concentration of collagen (1 mg/mL). The cells were subsequently incubated at 37°C for 72 hours. Upon the completion of the growth phase, the morphological properties of the cells were studied. Observations under bright-field microscopy revealed the directional alignment of HSCs grown on LC collagen films in comparison to the control substrate, as evidenced by Fig. 3A and C. Fluorescence imaging further confirmed the directional preference of HSCs when cultured on LC collagen films as opposed to the control sample (shown in Fig. 3B and D).

The order parameter of HSC alignment was determined by ImageJ software, using equation (1), where θ represented the angle between the orientation of the long axis of each cell and the average direction of orientation of cholesteric bands. The order parameter was found to be 0.9 (as shown in Fig. 3E). This level of alignment has been regarded as highly ordered in previous reports [39]. Interestingly, the order parameter of HSCs was slightly elevated in comparison to that of the cholesteric bands themselves- which could be attributed to the tendency of cells to align with their neighbouring counterparts by adhering to the fibronectin network released by the adjacent cells [40]. The order parameter of HSCs cultured on control samples with unaligned collagen was determined to be approximately 0.2. The disparity between the order parameter of cells cultured on control substrate and those cultured on aligned collagen film emphasizes the importance of the emergent directional signals to HSCs via aligned cholesteric bands to achieve unidirectional orientation of cells.

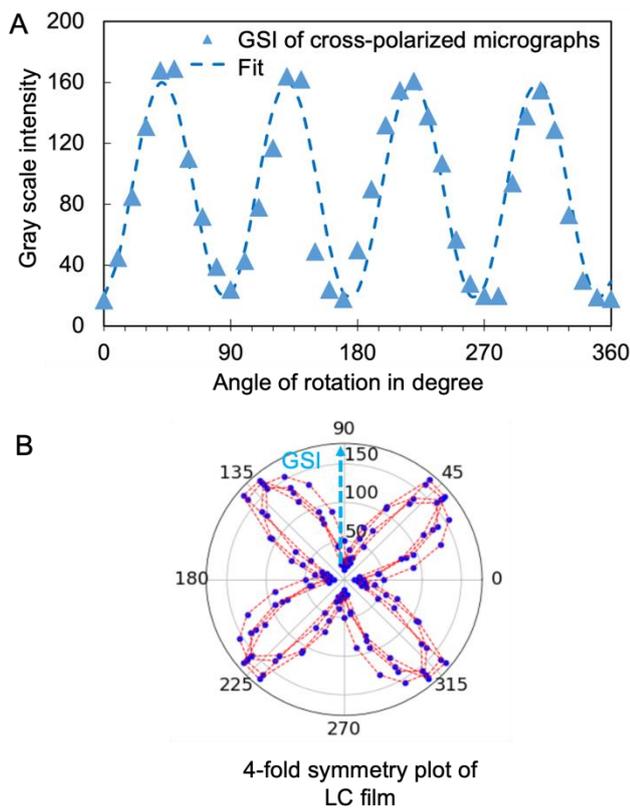

**Figure 2.** (A) The average intensity of grayscale value of a sequence of images vs. rotation angle (°). (B) Polar plot of gray scale intensity as a function of rotation angle (°), generated from a sequence of images taken from LC collagen film under cross-polarized light while rotating the microscope stage in 10° increments. The radial axis represents the gray scale intensity (GSI). The fitted equation is $Acos^2(B\theta + C) + D$.

In addition to the orientation of the HSCs, we quantified the aspect ratio (AR, the ratio of the major axis to the minor axis) of HSCs- optical micrographs clearly showed the qualitative shape change of the cells grown on aligned collagen when compared to cells grown on control substrates. AR was quantitatively evaluated utilizing the ImageJ software, for a minimum of 100 cells per statistical evaluation (as depicted in Fig. 3F). We find a significant enhancement in the AR of HSCs cultured on LC films, as compared to the control substrates, with the values of 4.2 and 1.7, respectively. While the exact mechanism of the shape change of the cells is beyond the scope of this work, it points to an exciting possibility of sharing of strain between the collagen and cells which has wide scale implications in understanding cellular properties. The sharing of strain is likely brought about by the distortion energy related to the disruption in the orientation of collagen due to the presence of an inclusion, HSCs in this instance. If the strain is coupled to the cellular membrane, the collagen can release some of the strain by stretching the cell membrane [29].

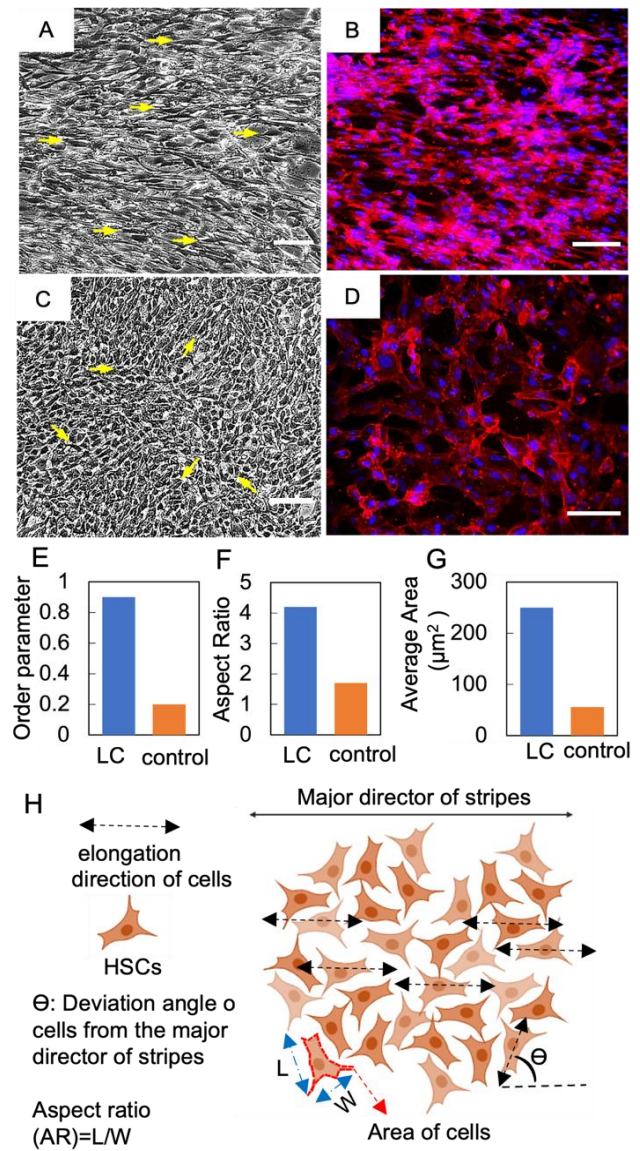

**Figure 3.** Comparison of HSCs behavior on LC and randomly oriented (serving as a control substrate) collagen films (A) and (B) Bright field and fluorescent image of HSCs cultured on LC collagen film, respectively. (C) and (D) Bright field and fluorescent image of HSCs, respectively, cultured on the control substrate. Areas marked in blue, and red are the nuclei and actin, which are labeled with Hoechst and Actin Red, respectively. (E), (F) and (G) Comparison of the order parameter, aspect ratio, and average area of HSCs, respectively, cultured on the LC and control sample. (H) Illustration of the angle of deviation between the elongation axis of HSCs and the main direction of cholesteric bands. The order parameter and aspect ratio of the cells were calculated by averaging over the entire population in each frame. To measure the average cell area, a thresholding algorithm was used to segment cells from the background in ImageJ. The scale bars represent 100μm.

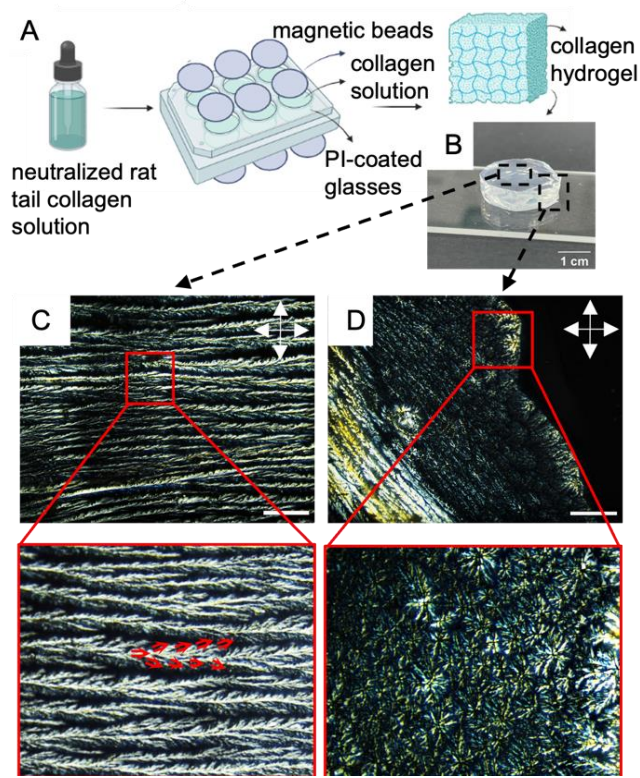

**Figure 4.** Formation and characterization of liquid crystalline collagen hydrogel under applied magnetic field: (A) Illustration of a 6-well plate setup in which rubbed polyimide (PI)-coated glasses are placed inside the wells and covered with a neutralized collagen solution (10mg/ml). Subsequently, the setup is subjected to an applied magnetic field and left to crosslink for 2 hours at 37°C. (B) Visual appearance of the liquid crystalline collagen hydrogel as observed with the unaided eye. (C) and (D) Optical micrographs of the collagen hydrogel, in bulk and edge, respectively, captured using cross-polarized light. Edge-induced disclinations composed of centrically rearranged cholesteric bands is visible in (D). The scale bars represent 30µm. The double-sided arrows indicate the direction of analyzer and polarizer.

HSCs also displayed a significantly increased spreading when cultured on collagen films compared to the control sample (Fig. 3G). The ability of cells to spread is a crucial aspect of cellular behaviour, as the inhibition of cell spreading triggers apoptosis [41-43]. Thus, understanding the effect of ECM anisotropy on cell spreading is of significance. The average area of HSCs cultured on both liquid crystalline (LC) collagen films and the control sample was quantitatively assessed using ImageJ software. The results demonstrated that the average area of HSCs was significantly larger on LC collagen films in comparison to the control substrate, indicating an enhancement in cellular spreading on the anisotropic film. To determine the average cell area, the brightfield microscopy images were converted to 8-bit images using ImageJ. Subsequently, individual cells were segmented from the background using a thresholding algorithm. The binary image was then subjected to a water shedding procedure, which separated the clustered cells. The area of the cells was then calculated using the "analyze particles" function on ImageJ. This approach enabled precise quantification of the cell area, which is essential for further downstream analysis of cellular properties. The average area of HSCs cultured on LC film and control substrate was measured as 250 µm$^2$ and 56 µm$^2$, respectively. Thus, it was interesting to see that not only the aspect ratio, but also the average area of HSCs can be regulated by the spatial organization of collagen fibrils.

## 2.3. Preparation and characterization of large-scale, monodomain 3-D liquid crystalline collagen hydrogels

3-D replication of the native environment is critical to addressing several biological questions as opposed to the simplistic version of using 2-D substrates. Therefore, we subsequently focused on making liquid crystalline collagen hydrogels (LCCHs) by incubating a solution of neutralized concentrated collagen (final concentration of 15 mg/mL), 20% PBS 1X, 0.1 wt.% glutaraldehyde, and 1 µg/mL fibronectin on rubbed PI-coated glasses in 6-well plates, while being subjected to a magnetic field with the direction of magnetic field perpendicular to the plane of the glasses (shown in Fig. 4A). The incubation was carried out at 37°C overnight to promote gelation and crosslinking.

In Fig. 4B, a typical disk-shaped LCCH with a thickness of 5mm is presented as an example. Fig. 4C presents a cross-polarized optical micrograph of the surface of the LCCH, depicting parallel cholesteric bands with two prominent morphological features, namely tree-like branching and braided structure. The red arrows in Fig. 4C trace the progression of individual cholesteric bands branching into further distinct cholesteric bands and eventually leaving the plane of observation. The extinction of the birefringent cord along the height of the frame results in the braided-like appearance of the cholesteric bands. Fig. 4D illustrates the occurrence of defects with concentrically arranged cholesteric bands on the edge of the hydrogel. These defects arise from the nucleation of disclinations, which is triggered by the curvature-induced stresses present in the hydrogels. The occurrence of these defects can likely be prevented by utilizing larger magnets, wherein the size of the hydrogel is significantly smaller relative to the area of the magnetic field- ensuring a homogenous magnetic field distribution throughout the hydrogel.

To quantify the LC order within the hydrogels, we analyze the variations in interference colour as a function of angular orientation. Changes in the interference colour of LCCHs were measured under cross-polarized light as the microscope stage was rotated by 15º increments counter-clockwise (Fig. 5A), with the light intensity and polarizer-analyzer correspondence being fixed. The anisometry in the optical properties of the hydrogel is depicted in Fig. 5B and C via the 4-fold symmetry of gray scale intensity plot with respect to the rotation angle and the cos$^2$ fit versus the rotation angle; respectively. To verify the role of magnetic forces in the formation of aligned cholesteric bands in the hydrogels, the same precursor solution

ARTICLE

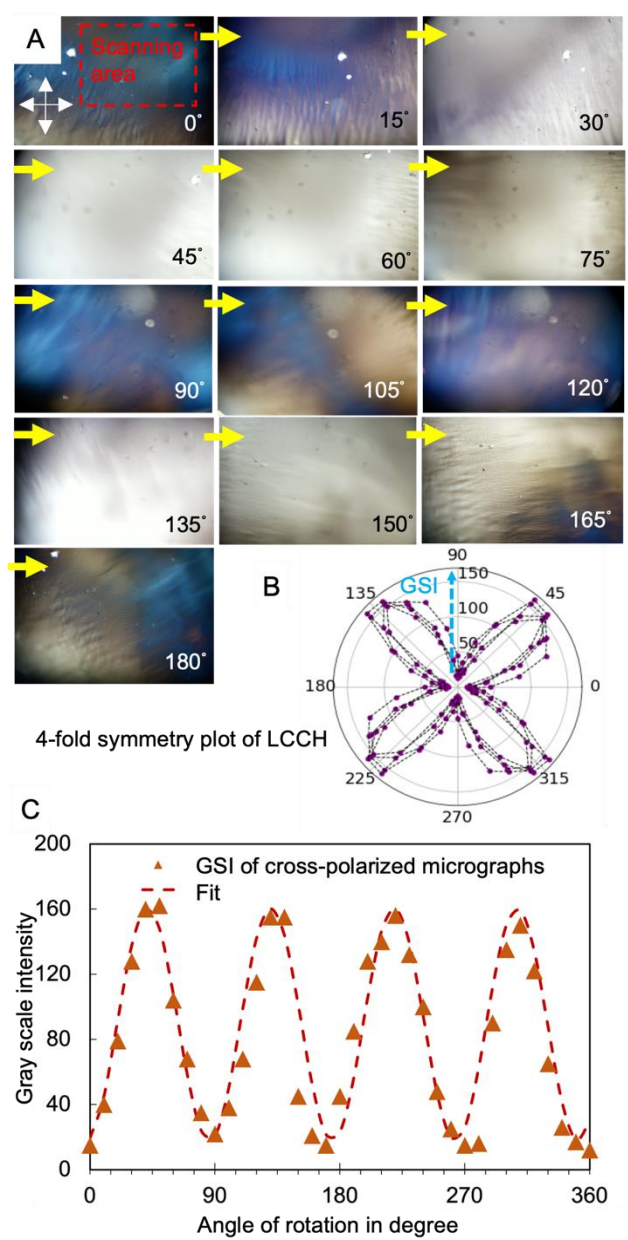

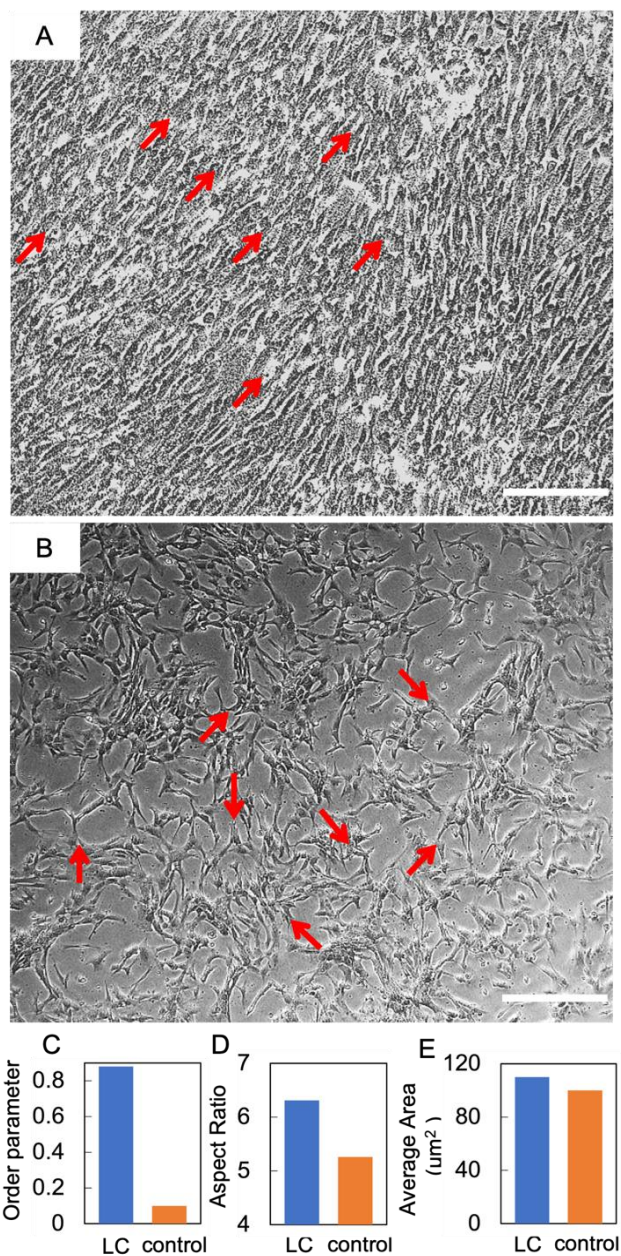

**Figure 5.** (A) Demonstration of range of interference color achieved for collagen hydrogel under cross-polarized light as the microscope platform rotates by 20° progressively in a clockwise manner. (B) Polar plot of gray scale intensity as a function of rotation angle (°), generated from a sequence of images taken from LC collagen hydrogel under cross-polarized light while rotating the microscope stage in 10° increments. The radial axis represents the gray scale intensity. Scale bar represents 100 µm. (C) The average intensity of grayscale value of a sequence of images vs. rotation angle (°). The fitted equation is $Acos^2(B\theta + C) + D$.

**Figure 6.** Comparison of HSCs behavior on LC and randomly oriented (serving as a control substrate) collagen hydrogels: (A) and (B) Bright field image of HSCs cultured on LC and control hydrogel, respectively. Scale bar represents 100µm. (C), (D) and (E) Comparison of HSCs behavior on LC and randomly oriented (serving as a control substrate) collagen hydrogels. Comparison of the order parameter, aspect ratio, and average area of HSCs, respectively, cultured on the LC and control collagen hydrogels.

of collagen was placed in a 6-well plate containing bare microscope glass, in the absence of a magnetic field, until a gel structure was formed. The complete extinction of light under the cross-polarized optical microscope for this control test (Fig. S3) demonstrates the isotropic structure of collagen fibrils in the resulted hydrogel.

### 2.4. Characterization of the behaviour of HSCs cultured on LCCHs

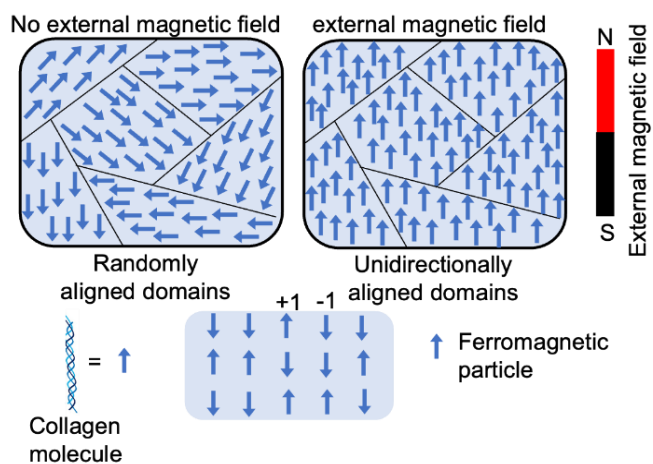

**Figure 7.** Schematic depiction elucidating the alignment of ferromagnetic particles influenced by an external magnetic field. Favorable contributions to the Boltzmann energy distribution in the Monte Carlo Simulation of Ising Theory (applied for the assembly of ferromagnetic particles) occur when spins align in the same direction (+1 or -1). collagen fibrils are treated as analogous entities to spins in the model.

LCCHs were positioned in 6-well culture plates and subjected to a culture medium comprised of HSCs and a low concentration of collagen (1 mg/mL). The cells were subsequently incubated at 37°C for a period of 72 hours. After the growth phase, the morphological properties of the cells were analyzed. Fig. 6A and B provide evidence that HSCs grown on LCCHs also exhibited directional alignment when compared to the control substrate.

The cells cultured on LCCHs exhibited an order parameter of 0.88, whereas those cultured on the controlled collagen hydrogels displayed a value of 0.1 (Fig. 6C), as determined by ImageJ analysis. Fig. 6D and E show the comparison of the aspect ratio and average area for cells cultured on LCCHs and control hydrogels.

### 2.5. Prediction of order parameter of LCCHs using Monte Carlo-Metropolis algorithm of Ising model

Collagen molecules possess diamagnetic properties which can be used to manipulate the alignment of collagen fibers [44, 45]. We model the alignment of the collagen fibers under a magnetic field using the Ising model for ferromagnetic particles as a simplistic model to capture some of the essential features of our observations- namely the ability to align macroscopic domains of collagen at modest magnetic field strengths. The Ising model without a magnetic field was first solved analytically by Lars Onsager in 1944 [46], and it has since been widely used to predict collective behaviour in a variety of fields. In the presence of a magnetic field, the Ising model can be numerically solved using the Metropolis algorithm, which is based on a few assumptions, including the fact that spins can either spin up or down with a tendency to align in the same direction as their neighbouring spins (Fig. 7).

To measure the distribution of probability for each state, the Boltzmann distribution with the partition function was used (equations (2) and (3)). At equilibrium, the probability of the system being in a certain energy configuration multiplied by the probability of that state transitioning to another state, summed over all possible states, is equal to the probability of the system being in another energy state multiplied by the probability of transitioning to that state, summed over all possible states (equation (4)). Numerically implementing the true-over-sum term in equation (4) can be challenging. Therefore, we can assume that it is valid for each individual term and enforce this by utilizing equation (5). To initiate the directional configuration, a random lattice of spins was generated. The system is then subjected to a temperature bath, and the equilibrium state of the orientation distribution of spins is determined numerically by converging to equilibrium state through a loop of flipping random spins and calculating the corresponding energy and deciding whether to keep that configuration via referencing to the ratio of the probability (equation (6)). The energy associated with each state is computed using equation (7) with a normalization factor J and summation over multiplication of spin directions for each 4 neighbouring spins. The system undergoes fluctuations governed by the Boltzmann equation until it reaches an equilibrium state.

$$P_\mu = \frac{1}{Z} e^{-\beta E_\mu} \quad (2)$$

$$Z = \sum_\mu e^{-\beta E_\mu} \quad (3)$$

$$\sum_\upsilon P_\mu P(\mu \to \upsilon) = \sum_\mu P_\upsilon P(\upsilon \to \mu) \quad (4)$$

$$P_\mu P(\mu \to \upsilon) = P_\upsilon P(\upsilon \to \mu) \quad (5)$$

$$\frac{P(\mu \to \upsilon)}{P(\upsilon \to \mu)} = \frac{P_\upsilon}{P_\mu} = e^{-\beta(E_\upsilon - E_\mu)} \quad (6)$$

$$E_\mu = \sum_{<i,j>} -J \sigma_i \sigma_j \quad (7)$$

At the equilibrium state, the average spin direction can be determined, which serves as an analogy to the order parameter of collagen fibrils. To account for the combined influence of collagen concentration and the strength of the external magnetic field, ß in equation (2) was substituted with a modified parameter (equation 8), where $H_c$ and $C_c$ represent the critical magnetic field strength and collagen concentration required for the formation of the LC phase.

$$\beta_m = \frac{\beta}{KT} \frac{H}{H_c} \frac{C}{C_c} \quad (8)$$

Fig. 8 presents the results derived from the computational analysis. Fig. 8A illustrates the evolution of the average spin direction over time for a reduced concentration ($C_R$) of 0.375 (where the critical concentration ($C_C$) is 80 mg/mL) and different reduced magnetic field strength ($H_R$) ranging from 0.2 to 0.9. The critical magnetic field ($H_C$) is set at 400 mT- this represents the minimum magnetic field strength we experimentally determined as necessary for the alignment of collagen for the concentration of 30 mg/mL. When $H_R$ is low (i.e., 0.2, 0.3, and 0.4), the average spin direction exhibits significant fluctuations, failing to converge to a steady state value over prolonged algorithmic iterations. Upon increasing $H_R$ to 0.5, however, these fluctuations diminish considerably, with the average spin direction exhibiting minor fluctuations around a constant value. For $H_R$ values exceeding 0.6, the average spin direction converges early in the algorithmic progression. Fig. 8B shows the absolute average spin direction (AASD) versus $H_R$ for various concentrations of collagen. An AASD of 0.4 is considered analogous to aligned cholesteric bands. The plot is divided into three distinct regions: Region 1, characterized by randomly aligned orientations with an AASD up to 0.2; Region 2, representing an aligned state with AASD ranging from 0.2 to 0.4; and Region 3, identifying a highly aligned configuration with AASD surpassing 0.4. Along constant $H_R$ (traceable by the vertical orange dashed line), as concentration ($C_R$) increases, AASD also rises, reflecting enhanced collaborative self-assembly of cholesteric bands with higher concentrations. Meanwhile, following the horizontal orange dotted line at constant AASD, an increase in $H_R$ results in a decrease in the required collagen concentration for the same AASD, highlighting the synergistic influence of magnetic strength and collagen concentration. For a constant $C_R$ (traceable by individual curves), elevating the magnetic strength increases the AASD, indicating an increase in the degree of alignment. The lag phase, representing the stage, where increasing the magnetic field minimally impacts AASD, varies with collagen concentration, for example, the lag phase is broader for $C_R$ of 0.2 compared to $C_R$ of 0.7 in Fig. 8B, emphasizing that there exists a minimum magnetic strength required for achieving alignment at each collagen concentration. Through this computational model, we reveal the intricate interplay between the collagen concentration and magnetic field strength, providing insights into the manipulation of its degree of alignment.

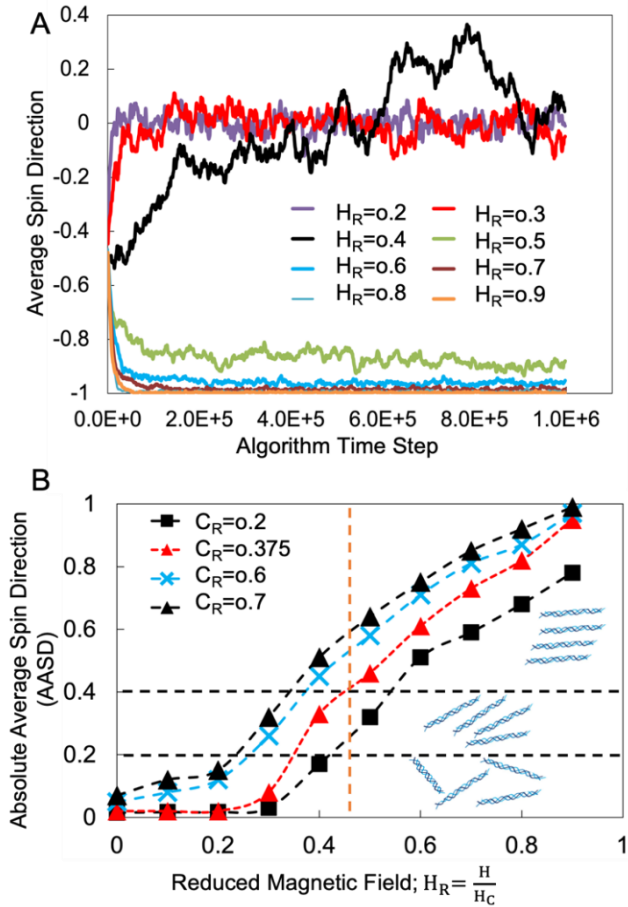

**Figure 8.** (A) Iterative convergence of spin directions under varying reduced magnetic fields for the reduced collagen concertation of 0.375. (B) Simulation-derived plot showing the absolute average spin direction (AASD) versus reduced magnetic field for various reduced concentration of collagen. The critical LC-forming collagen concentration is 80 mg/mL. The graph illustrates the progression from a randomly distributed state to a partially aligned intermediate phase and ultimately to a highly aligned configuration as the external magnetic strength increases.

## 3. Conclusion

Here, we harness the inherent anisotropic viscoelastic properties and diamagnetic susceptivity of collagen to create novel platforms composed of long-range unidirectionally aligned collagen (at concentrations well below the reported critical concentration of collagen to create LC phase). Both, films and hydrogels present with the anisotropy. While the alignment of HSCs along the direction of collagen orientation is intuitively expected, our results also show that, remarkably, the aspect ratio of HSC is a function of the alignment of collagen. Bright-field and fluorescent microscopy images demonstrate a higher aspect ratio, average area and order parameter in HSCs when cultured on LC platforms as compared to the control substrates. Our metropolis-based model of alignment of collagen provides guidance on magnetic field strength required to form aligned collagen substrates for a range of concentrations.

## 4. Materials and methods

### 4.1. Materials

The collagen type I, derived from rat tail, was obtained from Corning at a concentration of 7-11 mg/mL in 0.02N acetic acid. Prior to use, the collagen solution was concentrated by dialysis using 2K molecular weight cut-off (MWCO) Lyzer cassettes obtained from VWR, against an aqueous solution containing polyethylene glycol. Acetic acid and sodium hydroxide were procured from Sigma Aldrich, while the polyamide and PEG35k were sourced from HD Microsystems and VWR, respectively. Microscope slides were purchased from Ted Pella Inc. Fibronectin and glutaraldehyde were obtained from Sigma Aldrich. PBS 1X was purchased from Thermo Fisher Scientific.

### 4.2. Preparation of concentrated collagen

A stock mixture of collagen type I at a concentration of 7-11 mg/mL in 0.02N acetic acid was concentrated to prepare solutions with concentrations ranging from 15 to 40 mg/mL. The concentration procedure involved dialyzing the stock collagen solution against various concentrations of PEG35K (35-50 w/v%) in 150 mM acetic acid using 2k MWCO Lyzer cassettes. As the cassette only permitted water to pass through, the ultimate collagen concentration was calculated by deducing the volume of removed water from the initial volume, with the duration of the dialysis serving as the determinant factor.

### 4.3. Preparation of polyamide (PI)-coated substrates

Microscope slides with a thickness of 0.5 mm underwent a thorough cleaning procedure involving sonication in Alconox solution for 15 minutes, followed by washing with deionized water and ethanol. Following this, the slides were dried using an air gun and baked at 150°C for 1 hour, after which they were allowed to cool to room temperature. A spin coating technique was then employed, using a 1:10 ratio of polyimide (PI) to thinner, with a spin speed of 1000 rpm for 10 seconds, followed by 3000 rpm for 30 seconds. The coated slides were subsequently baked at 250°C for 2 hours. The resulting coating thickness was ~100 nm.

### 4.4. Formation of liquid crystalline films

Collagen fibrils in films were aligned utilizing a shear flow-driven method. The alignment layer comprised of PI was manually rubbed in a unidirectional manner using a microfiber cotton cloth, thereby creating parallel guiding grooves. The PI-coated glasses were subsequently washed with isopropyl alcohol and affixed in place using double-sided tape. Next, 50 μL of concentrated collagen solution was placed in a linear manner perpendicular to the grooves and casted in the same direction as the rubbing using a doctor blade with a constant speed, resulting in films with a thickness of approximately 150 μm. The coated substrates were allowed to air dry overnight at room temperature. Birefringence properties of the coatings were measured using an Olympus CX60 optical microscope with cross-polarized light.

### 4.5. Fabrication of liquid crystalline hydrogels

The PI-coated substrates were placed in 6-well glass culture plates. To each well, a solution of neutralized concentrated collagen (final concentration of 15 mg/mL), 20% PBS 1X, 0.1 wt.% glutaraldehyde, and 1 μg/mL fibronectin was added. Two magnetic beads were placed on the top and bottom of the wells to induce the alignment of collagen fibrils in the direction perpendicular to the magnetic field, thereby promoting a parallel orientation with the grooves. The culture plates were then incubated at 37°C overnight to facilitate the gelation and crosslinking of the collagen.

### 4.6. Cell culturing Human Schwann cells (HSCs) on collagen substrates.

HSCs from ATCC were cultured in a complete culture medium solution prepared with Dulbecco's Modified Eagle's Medium from Sigma Aldrich (Cat. #D5648), supplemented with 10% fetal bovine serum from Gibco (Cat. #10-437-028), and 1% penicillin–streptomycin from Sigma-Aldrich (Cat. #P4333), along with sodium bicarbonate and sodium pyruvate. The cells were incubated in a humid incubator at 37 °C with an air atmosphere containing 5% $CO_2$.

### 4.7. Instruments

The morphology of collagen substrates was observed using Olympus CX60 microscope integrated with a rotating stage. Bright field microscopy as well as fluorescent microscopy were used to observe the morphology of HSCs upon culturing. Ossila spin coater was used to apply the polymer solution on the microscope slides.

## Conflicts of interest

No conflicts to declare.

## Acknowledgements

KN acknowledges funding from Arkansas Biosciences. LH acknowledges funding from National Cancer Institute (NCI) Transition Career Development Award (K22-CA237857).

ARTICLE